\documentclass[a4paper,11pt]{article}
\usepackage{jheppub} 
\usepackage{lineno}
\usepackage{graphicx}
\usepackage{lipsum}
\usepackage{slashed}
\usepackage{amsmath}
\usepackage[normalem]{ulem}
\usepackage{enumitem}\usepackage{mathtools}
\usepackage{bm}
\usepackage[all]{hypcap}
\usepackage{tikz}
\usepackage{tikz-feynman}
\usepackage{bbm}
\usepackage{xcolor}
\usepackage{subcaption}
\usepackage[font=footnotesize,labelfont=bf]{caption}
\usepackage{mathabx}
\usepackage{pifont}
\usepackage{multirow}
\usepackage{tablefootnote}

\newcommand{\cC}{\mathcal{C}}

\title{
New Physics at Tera-$Z$: Precision Renormalised
}

\author[a]{Lukas Allwicher,} 
\emailAdd{lukas.allwicher@physik.uzh.ch}

\author[b]{Matthew McCullough,} 
\emailAdd{matthew.mccullough@cern.ch}

\author[c]{and Sophie Renner}
\emailAdd{sophie.renner@glasgow.ac.uk}

\affiliation[a]{Physik-Institut, Universit\"at Z\"urich, 8057 Z\"urich, Switzerland}
\affiliation[b]{Theoretical Physics Department, CERN, 1211 Geneva, Switzerland}
\affiliation[c]{School of Physics and Astronomy, University of Glasgow, Glasgow G12 8QQ, United Kingdom}

\abstract{We study the power of a Tera-$Z$ run at FCC-ee for indirectly detecting or constraining heavy new physics. Our main finding is that nearly every new particle which matches at tree level to dimension-six operators of the Standard Model Effective Field Theory (SMEFT) affects electroweak precision observables (EWPOs) at either tree level or via one loop renormalisation group (RG) running. This is true almost regardless of the structure of couplings to the Standard Model; just a handful of exceptions are identified which can produce zeroes in the EWPO RG equations. Under simple flavour assumptions, we perform fits of each state to projected FCC-ee $Z$ pole measurements, showing that all scenarios can be tested at the TeV scale or better, with many projected exclusions reaching tens of TeV. Tera-$Z$ is argued to provide an almost inescapable probe of heavy new physics.}

\begin{document}
\preprint{CERN-TH-2024-133, ZU-TH 39/24
}

\maketitle
\flushbottom

\section{Introduction}
The particle physics community has converged around an $e^+ e^-$ Higgs factory as the ideal next step in high energy physics.  While the physics opportunities offered for advancing our understanding of the microscopic nature of the Higgs boson have been comprehensively investigated, circular $e^+ e^-$ Higgs factories also offer the prospect of a Tera-$Z$ programme which has, in contrast, received much less dedicated attention.

The unprecedented microscopic resolution offered by trillions of $Z$ bosons recorded in a clean environment would unveil the interplay between the electroweak (EW), Higgs, QCD and flavour sectors of the Standard Model in a manner which cannot be understood from a na\"ive extrapolation of the LEP-I programme:  A view that would cast a Tera-$Z$ programme in terms of almost one million LEP-I programmes is shortsighted.

The inescapable interplay between all sectors of the SM arises due to the enormous improvement in statistical and systematic errors offered.  SM predictions would require the consideration of up to 3-loop EW processes.  With this many $\hbar$'s at play many corners of the SM play a role, virtually, in a single physical observable.  So too of physics beyond the SM (BSM).  While we are accustomed to viewing BSM physics through the lens of the specific sector, or fundamental question a given scenario is concerned with, such as the hierarchy or flavour puzzles, when it comes to mapping to observables such a tidy compartmentalisation would be impossible in a Tera-$Z$ era.

Looking forward to such a prospective era, if one were to ask the question `What BSM territory is explored by a given facility?' then the answer would be illustrated by the landscape of BSM possibilities for which that facility has a quantifiable potential for discovery.  Quantifying this potential is thus a challenging task for a Tera-$Z$ facility since accounting for quantum effects which may drag BSM effects into electroweak precision observables (EWPOs) is a non-negotiable requisite, even if the BSM scenario is not primarily concerned with the EW sector.

In some studies these effects have been accounted for (see e.g.\ \cite{Allwicher:2023shc,Stefanek:2024kds}), however thus far only in the context of specific model scenarios, or classes of scenarios; until now no attempt at a comprehensive mapping of BSM models into EWPOs for Tera-$Z$ has been attempted.  This work makes a first attempt in a top-down approach which ultimately maps BSM models to the SM Effective Field Theory (SMEFT) at the EW scale.

Our precise question is the following:  If some perturbative UV BSM scenario generates a tree-level contribution to a dimension-six SMEFT operator, does a Tera-$Z$ programme have sensitivity to this scenario at the one-loop level or better?  We find that the answer is \emph{yes}: there is sensitivity at one-loop or better to almost all perturbative UV models which generate a tree-level contribution to SMEFT at dimension six.  Since this holds for all perturbative UV completions, it is strongly suggestive that it would also hold for all strongly-interacting UV completions which generate contributions to SMEFT.

Specifically, the models considered are all of those which may generate dimension-six SMEFT operators at tree level, as catalogued, with tree-level matching coefficients, in \cite{deBlas:2017xtg} (the `Granada' catalogue). Future limits on these models, including from FCC-ee, have been studied without running in the recent global fits of Ref.~\cite{Celada:2024mcf}. The novel element considered here is that one-loop renormalisation group (RG) evolution from the new physics scale to the EW scale is accounted for, both in conceptual structural terms and in practical quantitative terms when estimating Tera-$Z$ reach. This RG-evolution brings nearly all models within Tera-$Z$ reach, even if the models don't contribute to EWPOs at tree level.  The RG contribution is IR-calculable and difficult to tune away, since it is proportional to a logarithm of a ratio of scales, unlike finite matching contributions.

Since they are highly model-dependent and not IR-calculable, one-loop finite matching terms are not included for all models, only in the isolated case where tree and RG contributions vanish.  While in general the finite matching terms may quantitatively influence the overall sensitivity they will not qualitatively change the overall picture since sensitivity is already present for all models in any case and removing sensitivity would require fine-tuning the one-loop matching terms against the IR-calculable RG contributions.

In Sec.~\ref{sec:observables} we outline the EWPOs considered and their mapping to SMEFT operators.  In Sec.~\ref{sec:exceptions} we describe the patterns of SMEFT operator running which can result in EWPO sensitivity, as well as potential exceptions which may escape EWPO constraints.  In Sec.~\ref{sec:fit} we present our quantitative estimates for model sensitivity through EWPOs at Tera-$Z$, demonstrating sensitivity for all cases.  We conclude in Sec.~\ref{sec:concs}.

\section{EWPOs and their EFT interpretation}
\label{sec:observables}

The traditional EWPOs are defined as combinations of partial decay widths of $Z$- and $W$-bosons and asymmetries, together with the $W$-boson mass, and the total decay widths $\Gamma_{Z,W}$ (see Table \ref{tab:ewobs} and \cite{Breso-Pla:2021qoe} for further details).
The $Z$ mass is canonically used as an input measurement to fix the electroweak parameters of the SM, so is not included in the set of EWPOs.
\begin{table}
    \centering
    \renewcommand{\arraystretch}{1.3}
    \begin{tabular}{|c|c|c|}
        \hline
         & Observable & Definition \\ \hline\hline
        \multirow{7}{*}{$Z$-pole} & $\Gamma_Z$ & $\sum_f \Gamma(Z\to f\bar f)$ \\
         & $\sigma_{\rm had}$ & $\frac{12\pi}{m_Z}\frac{\Gamma(Z\to e^+ e^-)\Gamma(Z\to q\bar q)}{\Gamma_Z^2}$ \\
         & $R_f$ ($f = e,\mu,\tau,c,b$) \tablefootnote{The $R_f$ ratios for leptons are usually defined in the inverse way, we put them together here for brevity.} & $\frac{\Gamma(Z\to f\bar f)}{\sum_q \Gamma(Z\to q\bar q)}$ \\
         & $A_f$ ($f = e,\mu,\tau,s,c,b$) & $\frac{\Gamma(Z\to f_L \bar f_L)-\Gamma(Z\to f_R \bar f_R)}{\Gamma(Z\to f\bar f)}$ \\
         & $A_{\rm FB}^{0,\ell}$ ($\ell=e,\mu,\tau$) & $\frac{3}{4} A_e A_\ell$ \\
         & $A_q^{\rm FB}$ ($q=c,b$) & $\frac{3}{4} A_e A_q$ \\
         & $R_{uc}$ & $\frac{\Gamma(Z\to u\bar u) + \Gamma(Z\to c\bar c)}{2\sum_q \Gamma(Z\to q\bar q)}$ \\ \hline
        \multirow{4}{*}{$W$-pole} & $m_W$ & \\
         & $\Gamma_W$ & $\sum_{f_1, f_2} \Gamma(W\to f_1 f_2)$ \\
         & ${\rm Br}(W \to \ell\nu)$ ($\ell = e,\mu,\tau$) & \\
         & $R_{W_c}$ & $\frac{\Gamma(W\to cs)}{\Gamma(W\to ud) + \Gamma(W\to cs)}$ \\ \hline
    \end{tabular}
    \caption{Definition of the EWPOs. For $W$ branching fractions to leptons both the individual branching fractions and universality ratios are measured (see \cite{Breso-Pla:2021qoe}).}
    \label{tab:ewobs}
\end{table}
Generic effects of heavy new physics (NP) in EWPOs are then usually parametrised as follows, in terms of modifications of the gauge boson-fermion couplings $\delta g$ and a $W$ boson mass shift $\delta m_W$:
\begin{align}
    \mathcal{L}_{\rm eff} \supset &- \frac{g_L}{\sqrt{2}}W^{+\mu}\left[\bar u_L^i \gamma_{\mu} \left(V_{ij}+\delta g_{ij}^{Wq}\right)d_L^j + \bar \nu_L^i \gamma_{\mu} \left(\delta_{ij}+\delta g_{ij}^{W \ell }\right)e_L^j\right]+{\rm h.c.}\nonumber \\
&-\sqrt{g_L^2+g_Y^2}\, Z^{\mu} \left( \bar f^i_L \gamma_{\mu}
\left( g_L^{Zf}\delta_{ij}+ \delta g_{L\,ij}^{Zf}\right) f_L^j
+\bar f^i_R \gamma_{\mu}
\left[ g_R^{Zf} \delta_{ij}+ \delta g_{R\,ij}^{Zf}\right] f_R^j
\right)\nonumber \\
&+\frac{g_L^2 v^2}{4}(1+\delta m_W)^2 W^{+\mu} W^-_{\mu}+\frac{g_L^2 v^2}{8 c_W^2}Z^{\mu} Z_{\mu}.
\label{EWEffLag}
\end{align}
There is no reason a priori that NP contributions should be flavour conserving. 
However, given the current experimental precision on electroweak observables, one expects the NP modifications to be subleading compared to the SM couplings.
The leading effects then come from the interference term between SM and NP, which will depend only on the flavour conserving pieces.\footnote{For the same reason, here we also neglect possible right-handed $W$ couplings, which can be generated by the SMEFT operator $\mathcal{O}_{Hud}$.}
Furthermore, in the dimension-six SMEFT one can make use of $SU(2)_L$ symmetry to relate $Z$- and $W$-boson coupling modifications.
Taking all this into account, the EW fit is sensitive to 20 parameters, which can be chosen to be \cite{Breso-Pla:2021qoe}
\begin{align}
    \delta g  \in & \{\delta g_{L\,11(22,33)}^{Z\nu}, \delta g_{L\,11(22,33)}^{Ze}, \delta g_{R\,11(22,33)}^{Ze}, \\
     &  \delta g_{L\,11(22)}^{Zu},\delta g_{R\,11(22)}^{Zu}, \delta g_{L\,11(22,33)}^{Zd}, \delta g_{R\, 11(22,33)}^{Zd}, \delta m_W\} \,.
\end{align} 
In \cite{Breso-Pla:2021qoe} it was shown that, given the lack of data with light quarks in the final states from LEP, the fit to traditional EWPOs exhibits one flat direction, which can however be lifted by considering forward-backward asymmetry measurements from the LHC.  For what follows we will therefore assume that all $\delta g$'s can be individually constrained.

\subsection*{EWPOs in SMEFT at tree level}
From a UV perspective, it is more useful to work in terms of SMEFT coefficients rather than $\delta g$'s, given that matching and running computations are available and partially automated in that language.
Matching the $\delta g$'s at tree level to dimension-six operators gives \cite{Allwicher:2023aql}
\begin{align}
    \delta g_{L\,ii}^{Z\nu}=&-\frac{v^2}{2}\left([\cC_{Hl}^{(1)}]_{ii}-[\cC_{Hl}^{(3)}]_{ii}\right)+\delta^U(1/2,0)\,,\\
    \delta g_{L\,ii}^{Ze}=&-\frac{v^2}{2}\left([\cC_{Hl}^{(1)}]_{ii}+[\cC_{Hl}^{(3)}]_{ii}\right)+\delta^U(-1/2,-1),\\
    \delta g_{R\,ii}^{Ze}=&\,-\frac{v^2}{2}[\cC_{He}]_{ii}+\delta^U(0,-1),\\
    \delta g_{L\,ii}^{Zu}=&\,-\frac{v^2}{2} \left([\cC_{Hq}^{(1)}]_{ii}-[\cC_{Hq}^{(3)}]_{ii}\right)+\delta^U(1/2,2/3),\\
    \delta g_{R\,ii}^{Zu}=&\,-\frac{v^2}{2}[\cC_{Hu}]_{ii}+\delta^U(0,2/3),\\
    \delta g_{L\,ii}^{Zd}=&-\frac{v^2}{2}\left([\cC_{Hq}^{(1)}]_{ii}+[\cC_{Hq}^{(3)}]_{ii}\right)+\delta^U(-1/2,-1/3),\label{eq:ZtobLbL}\\
    \delta g_{R\, ii}^{Zd}=&\,-\frac{v^2}{2}[\cC_{Hd}]_{ii}+\delta^U(0,-1/3),
\end{align}
where $\delta^U(T^3,Q)$ is
\begin{align}
\delta^U(T^3,Q)=&
-v^2\left( T^3+Q\frac{g_Y^2}{g_L^2-g_Y^2} \right)
\left(\frac{1}{4}\cC_{HD}+\frac{1}{2}[\cC^{(3)}_{H\ell}]_{22}+\frac{1}{2}[\cC^{(3)}_{H\ell}]_{11}-\frac{1}{4}[\cC_{\ell\ell}]_{1221}\right)
\nonumber\\
&-v^2Q\frac{g_Lg_Y}{g_L^2-g_Y^2} \cC_{HWB},
\end{align}
and we have approximated the CKM matrix as the identity.  One thus finds that the set of SMEFT coefficients entering at tree level in the EW observables are:
\begin{align}
\label{eq:EWPOops}
    \vec{\cC}_{\rm ew} = ([\cC_{H\ell}^{(1,3)}]_{ii},[\cC_{Hq}^{(1,3)}]_{ii},[\cC_{He}]_{ii},[\cC_{Hu}]_{ii},[\cC_{Hd}]_{ii},\cC_{HD},\cC_{HWB},[\cC_{\ell\ell}]_{1221}) \,,  
\end{align}
where $i=1,2,3$ except for $\cC_{Hu}$, for which the third-generation coefficient does not contribute to EWPOs.
This yields a total of 23 parameters. The  $\delta^U(T^3,Q)$ terms originate from SMEFT contributions to the input observables $\{m_Z, G_F, \alpha_{em}\}$ used to fix the parameters of the theory (see e.g.~Refs.~\cite{Han:2004az,Berthier:2015oma}).  For example, the operators $[\mathcal{O}^{(3)}_{H\ell}]_{11}$, $[\mathcal{O}^{(3)}_{H\ell}]_{22}$ and $\mathcal{O}_{1221}$ mediate $\mu\to e\bar \nu\nu$ and thus enter the measurement of $G_F$. Meanwhile, $\mathcal{O}_{HD}$ and $\mathcal{O}_{HWB}$ produce new contributions to $m_Z$ when the Higgs takes its vev. The coefficients of these operators hence have knock-on effects in all electroweak processes.

As can be seen from the matching conditions of the $\delta g$'s to SMEFT, the EWPOs may only probe linear combinations of SMEFT coefficients, with possible flat directions appearing in the fit in terms of $\vec{\cC}_{\rm ew}$.
In terms of parameter counting, we know that the observables are enough to individually constrain the 20 $\delta g$'s, i.e.~we expect three flat directions in the fit to the 23 SMEFT coefficients~\eqref{eq:EWPOops}.
Writing the matching schematically as
\begin{align}
    \delta g_i = A_{ij} \cC_j \,,
\end{align}
these flat directions can be identified with the Null Space of $A$.
It turns out that this space is spanned by the following linear combinations of coefficients:
\begin{align}
    \cC_0^{(1)} &\propto \,[\cC_{Hq}^{(1)}]_{33} - [\cC_{Hq}^{(3)}]_{33} , \label{eq:universal1}\\
    \cC_0^{(2)} &\propto \,-\frac{g_Y}{g_L}\,\cC_{HWB} + \sum_{i=1}^3 \left( [\cC_{H\ell}^{(3)}]_{ii} + [\cC_{Hq}^{(3)}]_{ii}] \right), \label{eq:universal2}\\
    \cC_0^{(3)} &\propto \,2\,\cC_{HD} -\frac{1}{2}\frac{g_L}{g_Y} \cC_{HWB} + \sum_{\psi} \sum_{i} Y_\psi \cC_{H\psi} \,,\label{eq:universal3}
\end{align}
where the sum over $i$ in the last equation again runs over the first two generations for $\cC_{Hu}$, and $i=1,2,3$ otherwise.
$\cC_0^{(1)}$ corresponds to a modification of the $Z$-boson coupling to left-handed top quarks, which is not probed by EWPOs.
The directions $\cC_0^{(2)}$ and $\cC_0^{(3)}$ are equivalent by equations of motion to $\cC_{HW}$ or $\cC_{HB}$ plus redundant operator coefficients that contribute only to triple-gauge or gauge-Higgs vertices~\cite{Grojean:2006nn,Brivio:2017bnu}. These do not contribute to EWPOs, but the flat directions can be lifted by including e.g.~$e^+e^- \to W^+W^-$ observables in a combined fit~\cite{Grojean:2006nn,Berthier:2016tkq}. It should also be kept in mind that these flat directions are not eigenvectors of the SMEFT anomalous dimension matrix, and will in general run into operator directions constrained by EWPOs.

Overall, then, 20 independent parameters in the dimension-six SMEFT can be constrained at tree level by EWPOs~\cite{Breso-Pla:2021qoe}. This is clearly a tiny subset of the full 2499 ($B$ and $L$ conserving) parameters of the SMEFT at dimension six. 

The purely bosonic operator coefficients $\cC_{HWB}$ and $\cC_{HD}$ correspond (up to normalisations) to the traditional oblique parameters $S$ and $T$, respectively~\cite{Peskin:1991sw}. Information on new physics from EWPOs is often distilled into constraints just within this two-dimensional parameter space. But it is worth remembering that to do so requires making particular assumptions on the form of the new physics, namely that it produces effects in bosonic operators only, and/or in linear combinations of operators which are equivalent to bosonic operators by equations of motion (see e.g.~\cite{Wells:2015uba}). This generally requires assuming that the coefficients $\cC_{H\psi}$, which induce gauge boson vertex corrections, are suppressed relative to the purely bosonic operators (or else that they appear in combinations exactly proportional to SM gauge currents). But as we shall see more explicitly below, many simple NP models generate these vertex correction operators at a lower loop order than bosonic operators, so they should not generally be ignored. Furthermore, the $S$ and $T$ parameters on their own do not constitute an RG-invariant subset, instead mixing with a broader class of operators, many of which also enter EWPOs~\cite{Wells:2015cre}. So any electroweak analysis taking account of the effects of running, as ours does, must go beyond $S$ and $T$ and consider the full set of dimension-six operator coefficients $\vec{\cC}_{\rm ew}$~\eqref{eq:EWPOops}.

\subsection*{EWPOs in SMEFT with one-loop RGE}

One-loop electroweak, QCD and top Yukawa corrections to $Z$ and $W$ pole observables in the SMEFT have been computed in \cite{Dawson:2019clf,Liu:2022vgo,Bellafronte:2023amz}.
However, for our purposes, it will be sufficient to consider only the tree-level coefficients above, and account for one-loop contributions from operators outside this set by RG evolving down from a NP scale $\Lambda$ to the EW scale where the observables are defined.

To understand which subset of SMEFT coefficients have the potential to generate the $\vec{\cC}_{\rm ew}$ coefficients at one loop in RGEs, it is helpful to classify the operators in terms of the kinematic properties of their amplitudes (specifically, the total helicity $\sum h$ and number of legs $n$), and their flavour structure. We can then use the non-renormalization theorems of Refs.~\cite{Cheung:2015aba} and~\cite{Machado:2022ozb} to delineate the sets of operator coefficients connected to EWPOs by running. 

All the operators corresponding to the list \eqref{eq:EWPOops} generate tree level amplitudes with $(n,\sum h)=(4,0)$, except for $\mathcal{O}_{HWB}$ whose amplitudes have $(n,\sum h)=(4,2)$. By the arguments of Ref.~\cite{Cheung:2015aba}, all operators producing amplitudes with $(n,\sum h)=(4,0)$ can mix into each other, meaning all SMEFT operators of the form $H^4D^2$, $\psi^2 \bar \psi^2$, or $\psi \bar \psi H^2 D$ could in principle run into EWPO operators. In addition, all operators whose amplitudes have $(n,\sum h)=(4,2)$ or $(n,\sum h)=(3,3)$ could mix with other $(n,\sum h)=(4,2)$ operators. This gives all SMEFT operators of the form $X^3$, $\psi^4$, $X^2H^2$, or $\psi \bar \psi X H$ the potential to run into $\mathcal{O}_{HWB}$ (where $X$ is any gauge field strength). In fact, the only operator types which are forbidden by helicity arguments from running into the EWPO operators at one loop are the Yukawa-like operators $\bar \psi \psi H^3$, and the six-Higgs operator $H^6$. 

The flavour criteria, it turns out, are much more stringent. If we ignore the effects of small Yukawa couplings in the running (i.e.~neglecting all but $y_t$, a very good approximation), and work in the basis where the up-type Yukawa matrix is diagonal, then the anomalous dimension matrix block-diagonalises according to the flavour structure of the operators~\cite{Machado:2022ozb}. In particular, no operator coefficients with flavour off-diagonal indices can generate the flavour diagonal (or purely bosonic) EWPO coefficients~\eqref{eq:EWPOops}. This drastically cuts down the parameter space, and we find that only 168 operator coefficients have the right flavour structure to run into EWPOs. This counting is true at any loop order, since it is based on the flavour symmetry structure of SMEFT operators and SM interactions.\footnote{To be specific, this counting includes: all CP-even flavour-singlet operators, all CP-even flavour-diagonal 4-fermion operators including a flavour-singlet current or a current with the flavour charges of $y_t^2$, the CP-even coefficients of $C_{uW}^{33}$, $C_{uB}^{33}$, $C_{uG}^{33}$, $C_{uH}^{33}$, and two extra flavour-diagonal components of $C_{ll}$. For details of the flavour charges of the SMEFT operators, see~\cite{Machado:2022ozb}.} This is only $\sim 7\%$ of the SMEFT parameter space at dimension six. 

\begin{table}[h]
  \begin{center}
    {\small
      \begin{tabular}{l|cccccccc}
        Scalar &
        ${\cal S}$ &
        ${\cal S}_1$ &
        ${\cal S}_2$ &
        $\varphi$ &
        $\Xi$ &
        $\Xi_1$ &
        $\Theta_1$ &
        $\Theta_3$ \\
         &
        $\left(1,1\right)_0$ &
        $\left(1,1\right)_1$ &
        $\left(1,1\right)_2$ &
        $\left(1,2\right)_{\frac 12}$ &
        $\left(1,3\right)_0$ &
        $\left(1,3\right)_1$ &
        $\left(1,4\right)_{\frac 12}$ &
        $\left(1,4\right)_{\frac 32}$ \\[1.3mm]
        &&&&&&&\\[-0.4cm]
         &
        ${\omega}_{1}$ &
        ${\omega}_{2}$ &
        ${\omega}_{4}$ &
        $\Pi_1$ &
        $\Pi_7$ &
        $\zeta$ &
        & \\ 
         &
        $\left(3,1\right)_{-\frac 13}$ &
        $\left(3,1\right)_{\frac 23}$ &
        $\left(3,1\right)_{-\frac 43}$ &
        $\left(3,2\right)_{\frac 16}$ &
        $\left(3,2\right)_{\frac 76}$ &
        $\left(3,3\right)_{-\frac 13}$ \\[1.3mm]
        &&&&&&&\\[-0.4cm]
         &
        $\Omega_{1}$ &
        $\Omega_{2}$ &
        $\Omega_{4}$ &
        $\Upsilon$ &
        $\Phi$ &
        &
        & \\
         &
        $\left(6,1\right)_{\frac 13}$ &
        $\left(6,1\right)_{-\frac 23}$ &
        $\left(6,1\right)_{\frac 43}$ &
        $\left(6,3\right)_{\frac 13}$ &
        $\left(8,2\right)_{\frac 12}$ \\[1.3mm]
\hline
Fermion &
        $N$ & $E$ & $\Delta_1$ & $\Delta_3$ & $\Sigma$ & $\Sigma_1$ & & \\
         &
        $\left(1, 1\right)_0$ &
        $\left(1, 1\right)_{-1}$ &
        $\left(1, 2\right)_{-\frac{1}{2}}$ &
        $\left(1, 2\right)_{-\frac{3}{2}}$ &
        $\left(1, 3\right)_0$ &
        $\left(1, 3\right)_{-1}$ & & \\[1.3mm]
        &&&&&&&\\[-0.4cm]
         &
        $U$ & $D$ & $Q_1$ & $Q_5$ & $Q_7$ & $T_1$ & $T_2$ \\
         &
        $\left(3, 1\right)_{\frac{2}{3}}$ &
        $\left(3, 1\right)_{-\frac{1}{3}}$ &
        $\left(3, 2\right)_{\frac{1}{6}}$ &
        $\left(3, 2\right)_{-\frac{5}{6}}$ &
        $\left(3, 2\right)_{\frac{7}{6}}$ &
        $\left(3, 3\right)_{-\frac{1}{3}}$ &
        $\left(3, 3\right)_{\frac{2}{3}}$ & \\[1.3mm]
\hline
        Vector &
        ${\cal B}$ &
        ${\cal B}_1$ &
        ${\cal W}$ &
        ${\cal W}_1$ &
        ${\cal G}$ &
        ${\cal G}_1$ &
        ${\cal H}$ &
        ${\cal L}_1$ \\
         &
        $\left(1,1\right)_0$ &
        $\left(1,1\right)_1$ &
        $\left(1,3\right)_0$ &
        $\left(1,3\right)_1$ &
        $\left(8,1\right)_0$ &
        $\left(8,1\right)_1$ &
        $\left(8,3\right)_{0}$ &
        $\left(1,2\right)_{\frac 12}$ \\[1.3mm]
        &&&&&&&\\[-0.4cm]
         &
        ${\cal L}_3$ &
        ${\cal U}_2$ &
        ${\cal U}_5$ &
        ${\cal Q}_1$ &
        ${\cal Q}_5$ &
        ${\cal X}$ &
        ${\cal Y}_1$ &
        ${\cal Y}_5$ \\
         &
        $\left(1,2\right)_{-\frac 32}$ &
        $\left(3,1\right)_{\frac 23}$ &
        $\left(3,1\right)_{\frac 53}$ &
        $\left(3,2\right)_{\frac 16}$ &
        $\left(3,2\right)_{-\frac 56}$ &
        $\left(3,3\right)_{\frac 23}$ &
        $\left(\bar 6,2\right)_{\frac 16}$ &
        $\left(\bar 6,2\right)_{-\frac 56}$ \\[1.3mm]
      \end{tabular}
    }
    \caption{Heavy scalars, fermions and vectors which can contribute to SMEFT at dimension-6, taken from Ref.~\cite{deBlas:2017xtg}, with corresponding representations under $\text{SU}(3)_C \times \text{SU}(2)_L \times \text{U}(1)_Y$.}
    \label{tab:Granadadict}
  \end{center}
\end{table}

\section{Categorising states that can escape EWPO constraints}
\label{sec:exceptions}
As argued in the previous section, EWPOs are sensitive to a rather small subset of SMEFT dimension-six parameters, even at loop level. But the SMEFT parameter space is not populated democratically by models of BSM physics; some operator directions are generated more easily or more inevitably by simple UV completions. So in the remainder of this paper, we focus on single-particle BSM extensions which match to the dimension-six SMEFT at tree level, as categorised in Ref.~\cite{deBlas:2017xtg}.  All such heavy states are detailed in Tab.~\ref{tab:Granadadict}.  We start in this section by studying these states in generality, investigating for each case whether it is possible to \emph{not} generate one or more of the operator coefficients in~\eqref{eq:EWPOops} at tree level or via one-loop RG running. This will provide an exhaustive overview of the potential of EWPOs to constrain each single-particle extension, and highlight any exceptions or loopholes that can be found by particular coupling choices. Having first undertaken this general survey, in the next section we perform fits to each state under simple coupling assumptions, showing the projected constraints from a future Tera-$Z$ run at FCC-ee.

\begin{figure}
    \centering
    \includegraphics[width=0.9\linewidth]{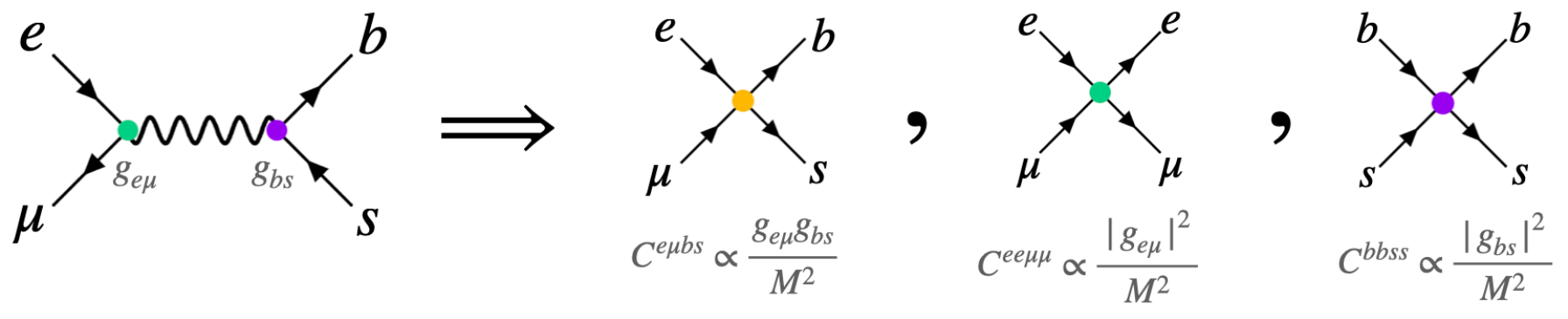}
    \caption{Schematic illustration showing that flavour violating operators are inevitably accompanied by flavour conserving operators within explicit UV completions. The tree level diagram on the left matches to the flavour violating 4-fermion coefficient $C^{e\mu b s}$, but by duplicating either one of the vertices, the flavour conserving coefficients $C^{e e\mu \mu}$ and $C^{bbss}$ are also generated from similar tree level diagrams.}
    \label{fig:FVUV}
\end{figure}

We have argued (based on Ref.~\cite{Machado:2022ozb}) that flavour off-diagonal operators are prevented from running into EWPOs, hence protecting much of the SMEFT parameter space from stringent electroweak precision tests, although in such instances flavour constraints would feature prominently. However, once we shift the focus onto tree-level mediators, it becomes clear that it is hard to generate flavour-violating coefficients without also generating flavour-conserving coefficients of the same magnitude or greater. This is illustrated schematically in Fig.~\ref{fig:FVUV}. More model-independent arguments can be made using sum rules~\cite{Remmen:2020uze}, demonstrating that flavour-violating operators are bounded above by flavour-conserving operators in a broad class of UV theories.

For this reason, we will find (up to a handful of exceptions) that any mediator which matches at tree level onto any part of the dimension-six SMEFT will generate EWPOs either at tree level or via one-loop running. Flavour violation in the couplings does not usually change this conclusion.

\subsubsection*{EWPOs at tree-level}
For an extension to the SM to produce no net effect in EWPOs at tree level, it must not match at tree level to any of the 23 EWPO coefficients defined in Eqn.~\eqref{eq:EWPOops}. Exceptions to this requirement are theories which match only to a linear combination of the $\cC_0^{(i)}$ coefficients in Eqns.~\eqref{eq:universal1}, \eqref{eq:universal2} and \eqref{eq:universal3}. The only example of this among the single-particle BSM extensions is the vector-like quark $U$, which if it only couples to third-generation quark doublets matches exactly onto the linear combination~\eqref{eq:universal1}. 

\begin{table}[t]
\begin{tabular}{|c | c | c| c|}
\hline
State & Tree level zero possible? & 1-loop RGE zero possible? & Bounding obs.\\
\hline 
  ${\cal S}$ & \ding{51} & \ding{55} & $A_e$\\
       ${\cal S}_1$ & \ding{51} if $(y_{{\cal S}_1})_{12}=0$ & \ding{55} & $A_e$\\
        ${\cal S}_2$ &\ding{51} & \ding{55} & $A_e$ ($R_\tau$)\\
        $\varphi$ &\ding{51} & \ding{55} & $\Gamma_Z$ \\
        $\Xi$ &\ding{55} & \ding{55} & $m_W$\\
        $\Xi_1$ &\ding{51} if $\kappa_{\Xi_1}=0$ \& $(y_{\Xi_1})_{12}=0$ & \ding{55} & $m_W$\\
        $\Theta_1$ &\ding{51} & \ding{51}$^*$ & $m_W$\\
        $\Theta_3$ &\ding{51} & \ding{51}$^*$ & $m_W$\\
        ${\omega}_{1}$ &\ding{51} & \ding{55} & $\Gamma_Z$\\
        ${\omega}_{2}$ &\ding{51} & \ding{55} & $\Gamma_Z$\\
        ${\omega}_{4}$ &\ding{51} & \ding{55} & $A_e$ ($R_\tau$)\\
        $\Pi_1$ &\ding{51} & \ding{55} & $A_e$ ($R_\tau$)\\
        $\Pi_7$ &\ding{51} & \ding{55} & $R_\tau$\\
        $\zeta$ &\ding{51} & \ding{55} & $\Gamma_Z$ ($R_\tau$)\\
        $\Omega_{1}$ &\ding{51} & \ding{55} & $\Gamma_Z$\\
        $\Omega_{2}$ &\ding{51} & \ding{55} & $\Gamma_Z$\\
        $\Omega_{4}$ &\ding{51} & \ding{51} if $(y_{\Omega_4})_{33}\neq 0$, all else zero  & $\Gamma_Z$ (-) \\
        $\Upsilon$ &\ding{51} & \ding{55} & $\Gamma_Z$\\
        $\Phi$ &\ding{51} & \ding{55} & $\Gamma_Z$\\
        \hline
\end{tabular}
\caption{Scalars, and whether they can escape EWPOs. Flavour indices are in the basis in which up-type Yukawas are diagonal. A \ding{55} means that at least one $Z$ pole operator will always be generated unless all couplings are zero. The asterisks signify that although all RGEs are zero, finite contributions to $\cC_{HD}$ are generated at one loop by $\Theta_1$ and $\Theta_3$~\cite{Durieux:2022hbu,Anisha:2021hgc}. The last column indicates the observable driving the bound in the case of flavour-universal (third generation only) couplings, without additional couplings being set to zero. Complementary information concerning the specific SMEFT operators generated is presented in App.~\ref{ap:ops}, Tab.~\ref{tab:ops_scalars}. \label{tab:scalarexceptions}}
\label{tab:scalars}
\end{table}

In Tables~\ref{tab:scalarexceptions}, \ref{tab:fermionexceptions} and \ref{tab:vectorexceptions}, we list the single-particle extensions of the SM alongside whether they contribute to EWPOs. A tick in the second column signifies that the state does not contribute to any of the EWPO operators at tree level, whereas a cross signifies that tree level EWPO contributions are unavoidable for that state if it generates any dimension-six SMEFT operators at tree level. Many of the scalar and vector BSM states only match to four-fermion operators at tree level, and therefore can produce zero effect in EWPOs at tree level. Some others \emph{can} have zero tree level matching to the EWPO operators if one of their couplings is zero; in particular, if a coupling to the Higgs can be set to zero then $\cC_{Hf}$ coefficients will not be generated. New fermions, on the other hand, will always match to one or more of the $\cC_{Hf}$ coefficients, but the couplings of $Q_1$ and $Q_7$ can be chosen such that only $\cC_{Hu}^{33}$ is generated, which is not an EWPO operator.

\begin{table}
\begin{tabular}{|l | c | c|c|}
\hline
State & Tree level zero possible? & 1-loop RGE zero possible? & Bounding obs.\\
\hline 
        $N$ & \ding{55} & \ding{55} & $A_e$ ($\Gamma_Z$)\\
        $E$ &  \ding{55} & \ding{55} & $\Gamma_Z$ ($R_\tau$)\\
        $\Delta_1$ & \ding{55} & \ding{55} & $A_e$ ($R_\tau$)\\
        $\Delta_3$ & \ding{55} & \ding{55} & $A_e$ ($R_\tau$)\\
        $\Sigma$ & \ding{55} & \ding{55} & $\Gamma_Z$ ($R_\tau$)\\
        $\Sigma_1$ & \ding{55} & \ding{55} & $A_e$ ($\Gamma_Z$)\\
       $U$ & \ding{51} if $(\lambda_U)_3 \neq 0$, all else zero & \ding{55} & $\Gamma_Z$ ($m_W$)\\
       $D$ & \ding{55} & \ding{55} & $\Gamma_Z$\\
       $Q_1$ & \ding{51} if $(\lambda_{Q_1}^u)_3\neq 0$, all else zero & \ding{55} & $m_W$\\
       $Q_5$ & \ding{55} & \ding{55} & $\Gamma_Z$\\
       $Q_7$ & \ding{51} if $(\lambda_{Q_7})_3\neq 0$, all else zero & \ding{55} & $m_W$\\
       $T_1$ & \ding{55} & \ding{55} & $m_W$ ($\Gamma_Z$)\\
       $T_2$ &\ding{55} & \ding{55} & $\Gamma_Z$\\
       \hline
\end{tabular}
\caption{Same as for Tab.~\ref{tab:scalarexceptions}, but for BSM fermions.
\label{tab:fermionexceptions}}
\label{tab:fermions}
\end{table}

\begin{table}
\begin{tabular}{|l | c | c| c|}
\hline
State & Tree level zero possible? & 1-loop RGE zero possible? & Bounding obs.\\
\hline 
        ${\cal B}$ &\ding{51} if $(g_{\mathcal{B}}^\phi)=0$ \& $(g_{\mathcal{B}}^l)_{12}=0$ & \ding{51}  if eqns.~\eqref{eq:RGEB} & $m_W$\\
        ${\cal B}_1$ &\ding{51} if $(g_{\mathcal{B}_1}^\phi)=0$  & \ding{55} & $m_W$\\
        ${\cal W}$ &\ding{51} if $(g_{\mathcal{W}}^\phi)=0$ \& $K_{12}(g_{\mathcal{W}}^l)=0$ & \ding{51} if eqns.~\eqref{eq:RGEWcond} & $A_e$ ($\Gamma_Z$)\\
        ${\cal W}_1$ &\ding{55} & \ding{55} & $m_W$\\
        ${\cal G}$ &\ding{51} & \ding{51} if $(g^u_{\cal G})_{33}\neq 0$, all else zero & $\Gamma_Z$\\
        ${\cal G}_1$ &\ding{51} & \ding{55} & $\Gamma_Z$\\
        ${\cal H}$ &\ding{51} & \ding{55} & $\Gamma_Z$\\
      ${\cal L}_1$ &\ding{55} & \ding{55} & $A_e$ \\
        ${\cal L}_3$ &\ding{51} & \ding{55} & $A_e$ ($A_\tau$)\\
        ${\cal U}_2$ &\ding{51} & \ding{55} & $\Gamma_Z$\\
        ${\cal U}_5$ &\ding{51} & \ding{55} & $R_\tau$\\
        ${\cal Q}_1$ &\ding{51} & \ding{55} & $R_\tau$\\
        ${\cal Q}_5$ &\ding{51} & \ding{55} & $\Gamma_Z$\\
        ${\cal X}$ &\ding{51} & \ding{55} & $R_\tau$ \\
        ${\cal Y}_1$ &\ding{51} & \ding{55} & $\Gamma_Z$\\
        ${\cal Y}_5$ &\ding{51} & \ding{55} & $\Gamma_Z$\\
\hline
\end{tabular}
\caption{Same as for Tab.~\ref{tab:scalarexceptions}, but for BSM vectors. For the case of $\mathcal{W}$ we define the relevant first- and second-generation coupling flavour combination as $K_{12}(g_{\mathcal{W}}^l) = 2 (g_{\mathcal{W}}^l)_{11} (g_{\mathcal{W}}^l)_{22}-(g_{\mathcal{W}}^l)^2_{12}$.
\label{tab:vectorexceptions}}
\label{tab:vectors}
\end{table}

\subsubsection*{EWPOs and one-loop RG}
While it is true that many scalars and vectors may not contribute to the EWPOs at tree level, given the expected precision it is also important to investigate whether any states can escape contributions to EWPOs at loop level. For an extension to the SM to produce zero effect in EWPO operators at one-loop in RGEs, it must \emph{additionally} produce zeroes in the right-hand sides of all of the following equations~\cite{Jenkins:2013zja,Jenkins:2013wua,Alonso:2013hga}:\footnote{We omit the RGE for the operator coefficient $\cC_{HWB}$ here, as it depends only on coefficients which are not generated at tree level by weakly coupled UV completions~\cite{deBlas:2017xtg,Einhorn:2013kja,Craig:2019wmo}, and hence does not imply additional constraints on the couplings of the BSM states.}\footnote{We include a term dependent on $\cC_{Hq}^{(1)33}$ in Eqn.~\eqref{eq:CHDRGE}. On its own, this coefficient induces $Z\to \bar b b$ and hence would already contribute at tree level to EWPOs. However, in combination with an equal and opposite contribution to $\cC_{Hq}^{(3)33}$ (which does not enter the $\cC_{HD}$ RGE) the $Z\bar b b$ coupling exactly cancels. So we include this term to cover this possibility.}
\begin{align}
\dot{\cC}_{HD}&=N_c \left(\frac{16}{3}Y_H Y_u g_1^2-8y_t^2 \right)\cC_{Hu}^{33}+N_c \left(\frac{32}{3}Y_H Y_q g_1^2+8y_t^2 \right)\cC_{Hq}^{(1)33}\nonumber \\
&+\frac{80}{3}g_1^2 Y_H^2 \cC_{H\Box},\label{eq:CHDRGE}\\
\dot{\cC}_{Hf_1}^{(1)jj} &= 2 N_c y_t^2 \left( -S_{f_1 u}\cC_{u f_1}^{33jj}+S_{f_1 q}\cC_{qf_1}^{33jj} \right), \label{eq:CHf1RGEYuk}\\
\dot{\cC}_{Hl}^{(3)jj}&=-2N_c y_t^2 \cC_{lq}^{(3)jj33},\label{eq:CHl3RGEYuk}\\
\dot{\cC}_{Hq}^{(3)jj}&= -2y_t^2\left( 2N_c \cC_{qq}^{(3)jj33}+\cC_{qq}^{(1)j33j}-\cC_{qq}^{(3)j33j}\right),\label{eq:CHq3RGEYuk}\\
\dot{\cC}_{Hf_1}^{(1)jj}&=\frac{4}{3}g_1^2 Y_h \sum_k\bigg[  Y_{e} S_{f_1 e} \cC_{f_1 e}^{jjkk} + 2 Y_{l} S_{f_1 l} \cC_{f_1 l}^{jjkk} \nonumber \\&+ N_c \left(Y_{u} S_{f_1 u} \cC_{f_1 u}^{jjkk} +Y_{d} S_{f_1 d} \cC_{f_1 d}^{jjkk}+ 2 Y_{q} S_{f_1 q} \cC_{f_1 q}^{jjkk} \right) \bigg], \label{eq:CHf1RGE}\\
\dot{\cC}_{Hl}^{(3)jj}&=\frac{2}{3}g_2^2\sum_k\left(\cC_{ll}^{jkkj}  +N_c  \cC_{lq}^{(3)jjkk}\right),\\
\dot{\cC}_{Hq}^{(3)jj}&= \frac{2}{3}g_2^2 \sum_k\left(\cC_{lq}^{(3)kkjj}+2 N_c \cC_{qq}^{(3)kkjj}+\cC_{qq}^{(1)jkkj}-\cC_{qq}^{(3)jkkj} \right),\label{eq:CHq3RGE}\\
\dot{\cC}_{ll}^{1221}&= \frac{2}{3}g_2^2 \bigg[ \cC_{ll}^{2222}+ \cC_{ll}^{2332}+\cC_{ll}^{1111}+ \cC_{ll}^{1331}+N_c \sum_k \left( \cC_{lq}^{(3)22kk}+\cC_{lq}^{(3)11kk}\right) \bigg],\label{eq:Cll1221RGE}
\end{align}
where $f_1, f_2$ label fermion species $f_1,f_2 \in \{q,u,d,l,e\}$, $S_{f_1 f_2}=1+\delta_{f_1 f_2}$ is a symmetry factor, $Y_H$ is the hypercharge of the Higgs doublet, and $Y_{f_1}$ is the hypercharge of fermion $f_1$. The first four equations involve Yukawa contributions, keeping only the top Yukawa, while the remaining four involve only gauge contributions. Flavour indices are defined in the basis where the lepton and up-type quark Yukawas are both diagonal. Where there is more than one option for the interpretation of $\cC_{f_1 f_2}$ in Eqns.~\eqref{eq:CHf1RGEYuk} and \eqref{eq:CHf1RGE}, e.g.~in the case of $\cC_{lq}^{(1)}$ and $\cC_{lq}^{(3)}$, the $(1)$ superscript is always implied.

One can imagine that there could be accidental cancellations between the terms in the equations above, such that a particle which contributes to several operator coefficients (or several flavour components of a coefficient) might produce accidental zeroes in the RGEs overall. In practice, however, this occurs very rarely. The reason for this is that in many cases the RGE equations reduce to a sum of contributions of definite sign, meaning that they can only be zero if all couplings are zero. 

\begin{figure}
    \centering
    \includegraphics[width=\textwidth]{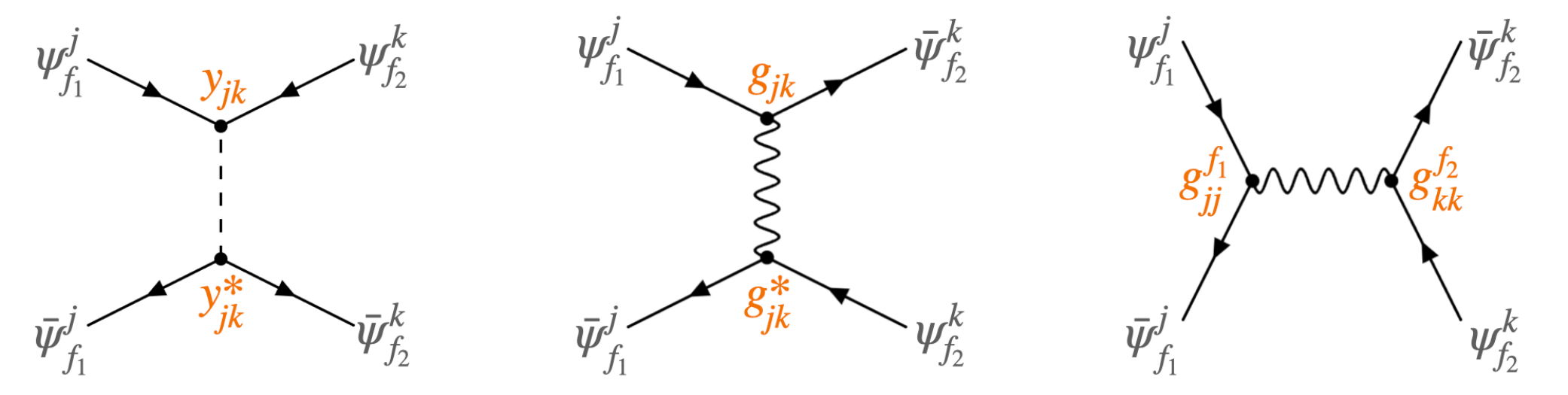}
    \caption{Diagrams illustrating the tree level matching of scalars and vectors onto $\cC_{f_1 f_2}^{jjkk}$. Directions of fermion arrows on the leftmost two diagrams will depend on the gauge charges of the mediator.}
    \label{fig:matchingdiagrams}
\end{figure}

To illustrate this, we can first consider the case of a scalar completion which only contributes to one four-fermion operator, through the diagram on the left in Fig.~\ref{fig:matchingdiagrams}. Then any resulting contributions to the gauge-dependent RGEs will be of the form:
\begin{equation}
    \label{eq:RGEdefsign}
    \dot \cC_{Hf_1}^{jj} \,\propto \sum_k |y_{jk}|^2,
\end{equation}
which can only be zero for all $j$ if all couplings $y_{jk}$ are zero. The same argument applies for any vector particle carrying hypercharge, which must therefore have couplings involving two different fermion species, as shown in the middle diagram of Fig.~\ref{fig:matchingdiagrams}. The resulting RGE contributions end up in a similar form as Eqn.~\eqref{eq:RGEdefsign}, with $y_{jk}\to g_{jk}$. 

For hypercharged particles which contribute to more than one four-fermion operator, the argument still applies. This can be easily seen in the case that the different operators have no fields in common and contribute to mutually exclusive RGEs, since the different couplings must then individually satisfy equations of the form Eqn.~\eqref{eq:RGEdefsign}. The other option is that the state matches to two operators which contain the same fields with different gauge contractions, for example $\cC_{lq}^{(1)}$ and $\cC_{lq}^{(3)}$. In this case the two coefficients are dependent on the same couplings, and one or more of the RGEs again reduce to the form in Eqn.~\eqref{eq:RGEdefsign}, which is only zero if all couplings are zero.

We therefore conclude that any scalar or vector particle which carries hypercharge and generates four-fermion operators \emph{must} contribute to EWPO operators either at tree level or via one-loop RGEs. The only possible exception to this is the scalar $\Omega_4$, which matches only to $\cC_{uu}$. If only its third generation coupling $(y_{\Omega_4})_{33}$ is non-zero, then it will generate only the coefficient $\cC_{Hu}^{33}$ at one loop in RGEs, which is not a $Z$ pole coefficient.\footnote{However, if the one-loop running is resummed, this coefficient does run into $C_{HD}$.} There are also two particles with hypercharge which do not generate four-fermion operators at tree level: $\Theta_1$ and $\Theta_3$ \cite{Logan:2015xpa,Chala:2018ari,deBlas:2014mba,Henning:2014wua,Jiang:2016czg,Dawson:2017vgm,Corbett:2017ieo,Murphy:2020rsh}. These therefore do not contribute to the RGEs above, but they do generate \emph{finite} one-loop contributions to $\cC_{HD}$, as calculated in Ref.~\cite{Durieux:2022hbu,Anisha:2021hgc}.

Having covered all states carrying hypercharge, we now turn to hypercharge-neutral states. There are two scalar particles with zero hypercharge: the $SU(2)_L$ singlet $\mathcal{S}$, and the $SU(2)_L$ triplet $\Xi$. The singlet $\mathcal{S}$ generates $\cC_{H\Box}$ at tree level via its trilinear coupling to the Higgs, which runs into $\cC_{HD}$ at one loop~\eqref{eq:CHDRGE}. It is not possible to forbid the matching of $\mathcal{S}$ to this operator without forbidding all of its tree level matching to dimension-six operators, which would take it outside of the scope of our analysis. The triplet $\Xi$ matches to the $Z$ pole operator $\cC_{HD}$ at tree level, and again it is not possible to forbid this coupling without eliminating all tree level matching.

Among the vector particles, there are four states with zero hypercharge ($\mathcal{B}$, $\mathcal{W}$, $\mathcal{G}$, $\mathcal{H}$) which couple to fermions.
In this case, their tree-level matching to four-fermion operators can go through the rightmost diagram in Fig.~\ref{fig:matchingdiagrams}, meaning that the resulting Wilson coefficients $\cC_{f_1f_2}^{jjkk}$ are of indefinite sign for each choice of $\{j,k\}$, and cancellations may be possible within RGEs. We now discuss each of these four vector states in turn.

The state $\mathcal{B}$ contributes to the right hand side of all the RGEs except~\eqref{eq:CHDRGE} and~\eqref{eq:CHl3RGEYuk} (assuming that its Higgs coupling $(g_{\mathcal{B}}^\phi)$ is zero; otherwise it contributes to EWPOs already at tree level). If we require these to be zero, we obtain the following conditions on the $\mathcal{B}$ couplings:
\begin{align}
    -\sum_{k=1}^3 \left((g_{\mathcal{B}}^e)_{kk}  + (g_{\mathcal{B}}^d)_{kk}\right) +2 \sum_{k=1}^2(g_{\mathcal{B}}^u)_{kk}&\overset{!}{=}0, ~~(g_{\mathcal{B}}^q)_{ij}\overset{!}{=}0, ~~ (g_{\mathcal{B}}^l)_{ij}\overset{!}{=}0, ~~  (g_{\mathcal{B}}^u)_{33}&\overset{!}{=}0.\label{eq:RGEB}
\end{align}
We can compare these to the charge assignments needed such that $\mathcal{B}$ can correspond to the gauge boson of an anomaly-free $U(1)$ gauge symmetry (see e.g.~Section 2 of Ref.~\cite{Allanach:2018vjg}). We find that anomaly-free scenarios can exist, in which the charges of the SM fermions are necessarily non-universal in flavour. For example, a viable anomaly-free charge assignment satisfying~\eqref{eq:RGEB} is $(g_{\mathcal{B}}^u)_{11}=(g_{\mathcal{B}}^d)_{11}=(g_{\mathcal{B}}^e)_{11}=-(g_{\mathcal{B}}^u)_{22}=-(g_{\mathcal{B}}^d)_{22}=-(g_{\mathcal{B}}^e)_{22}$, $(g_{\mathcal{B}}^u)_{33}=(g_{\mathcal{B}}^d)_{33}=(g_{\mathcal{B}}^e)_{33}=0$. It turns out that it is not possible to obtain anomaly-free charge assignments satisfying~\eqref{eq:RGEB} which are universal in the first two generations, and therefore any misalignment between the flavour and gauge bases will lead to tree level flavour changing neutral currents (FCNCs) among the light generations of quarks and/or leptons.\footnote{We thank Joe Davighi for providing a proof of this fact, and the viable anomaly-free charges we quote here.}

The state $\mathcal{W}$ matches to $\cC_{qq}^{(3)}$, $\cC_{ll}$ and $\cC_{lq}^{(3)}$ (if we take its coupling to the Higgs to be zero). Solving Eqns.~\eqref{eq:CHl3RGEYuk}--\eqref{eq:Cll1221RGE}, we find the coupling requirements to produce no contributions to the EWPO RGEs is: 
\begin{equation}
\label{eq:RGEWcond}
    (g^l_{\mathcal{W}})_{ij}\overset{!}{=}0,~~(g^q_{\mathcal{W}})_{3j}\overset{!}{=}0,~~(g^q_{\mathcal{W}})_{22}\overset{!}{=}\pm (g^q_{\mathcal{W}})_{11}, ~~ |(g^q_{\mathcal{W}})_{12}|\overset{!}{=}(g^q_{\mathcal{W}})_{11}\sqrt{4N_c-1}.
\end{equation}
Notice that this requires a large off-diagonal coupling involving the first two generations of quarks, and hence would induce large tree-level contributions to $K$ and $D$ meson mixing. 

The state $\mathcal{G}$ matches to all four-quark operators. Due to its octet colour structure, the matching is to $\cC_{uu}^{jjkk}\propto |(g_\mathcal{G}^u)_{jk}|^2$, and similar for $\cC_{dd}$. This means that the RGEs for $\cC_{Hu}$ and $\cC_{Hd}$ are in the form \eqref{eq:RGEdefsign}, meaning most couplings to $d$ and $u$ must be zero in order to not run into EWPOs at one loop. To see that there is no non-trivial solution involving the couplings to quark doublets requires a few more steps but can still be seen transparently. The matching gives $\cC_{qq}^{(1)ijkl}\propto (3 (g_\mathcal{G}^q)_{kj}(g_\mathcal{G}^q)_{il}-2 (g_\mathcal{G}^q)_{ij}(g_\mathcal{G}^q)_{kl})$ and $\cC_{qq}^{(3)ijkl}\propto 3 (g_\mathcal{G}^q)_{kj}(g_\mathcal{G}^q)_{il}$~\cite{deBlas:2017xtg}, where an equal prefactor has been removed from both coefficients. Then requiring a zero result for Eqn.~\eqref{eq:CHf1RGEYuk} for any choice of $j$ implies that $(g_\mathcal{G}^q)_{3j}\overset{!}{=}0$ for all $j$. Using this, additionally requiring a zero result for Eqn.~\eqref{eq:CHq3RGE} implies that $(g_\mathcal{G}^q)_{11}^2\overset{!}{=}(g_\mathcal{G}^q)_{22}^2\overset{!}{=}-|(g_\mathcal{G}^q)_{12}|^2$, implying that $(g^{\mathcal{G}})_{ij}\overset{!}{=}0$ for all $\{i,j\}$. Hence the only possibility for achieving an accidental zero in the one-loop $Z$ pole RGEs is $(g_\mathcal{G}^u)_{33}\neq 0$, all other couplings zero.

Finally, there is the state $\mathcal{H}$. This only matches at tree level to $\cC_{qq}^{(1,3)}$. In this case, again, there are no non-zero couplings which can cancel to give zero in the $Z$ pole RGEs. To see this, it is sufficient to notice that, given $\cC_{qq}^{(1)iijj}\propto |(g_\mathcal{H})_{ij}|^2$~\cite{deBlas:2017xtg}, Eqn.~\eqref{eq:CHf1RGE} is an equation of the form Eqn.~\eqref{eq:RGEdefsign}, so obtaining zero here implies that $(g^{\mathcal{H}})_{ij}\overset{!}{=}0$ for all $\{i,j\}$.

To summarize, all UV states which match at tree level to the dimension-six SMEFT (Tab.~\ref{tab:Granadadict}) can contribute to EWPOs at either tree level or one loop. However, in a handful of cases, specific coupling choices can be made which produce zeroes in the one-loop RGEs for the EWPO operators. It should be noted, however, that these zeroes are unlikely to persist under RG resummation. Also, the coupling choices required can imply large effects in other precisely-measured observables such as meson mixing.

\section{Probing one-particle extensions of the SM}
\label{sec:fit}
As discussed in the previous section, the majority of one-particle extensions of the SM give rise, either at tree or at loop level, to effects in EWPOs.
We now turn to quantitatively analysing the different options one by one, i.e. we consider one new state at a time, from the list of Ref.~\cite{deBlas:2017xtg}. In each case we study the projected constraints from a Tera-$Z$ run at FCC-ee, 
taking the sensitivities from Refs.~\cite{deBlas:2022ofj}.
We adopt three benchmark scenarios for the flavour structure of the couplings to the SM fermions:
\begin{itemize}
    \item[(a)] \textbf{Flavour-universal couplings}, i.e. $y_{11}=y_{22}=y_{33}$ (or $y_1 =y_2=y_3$ for fermions), and $y_{ij} = 0$ for $i\neq j$.\footnote{Note that in some cases (particularly for leptoquarks) this flavour scenario implies the tree level generation of flavour violating operators, and hence if this assumption is taken seriously, some states are not phenomenologically viable at the (tens of) TeV scale. However, flavour violating operators play no role at all in our analysis, thanks to the non-renormalisation theorems mentioned at the end of Section~\ref{sec:observables}. Furthermore, in more complete models there can be mechanisms to suppress these operators (e.g.~through flavour symmetries~\cite{Greljo:2023adz}) without affecting the flavour conserving operators which our study relies on. This flavour scenario can therefore be taken, for our purposes, as a condition on the flavour conserving operators only.} 
    \item[(b)] \textbf{Third-generation couplings only}, i.e. $y_{ij} = y \delta_{i3}\delta_{j3}$, which corresponds to a NP scenario with new states coupled only to the heaviest SM generation.
    \item[(c)] \textbf{Antisymmetric couplings}. Some of the scalar extensions considered are forced by gauge symmetry to have an antisymmetric coupling matrix in flavour space. For this reason, choosing either of the two options above would not be consistent, and in particular the third-generation only scenario is not defined. In order to show that these cases still retain sensitivity to EWPOs, we choose instead to switch on \emph{all} couplings, such that flavour-conserving operators are generated by the product of two flavour-violating couplings. This is the case for $S_1$ and $\omega_2$ in Fig. \ref{tab:scalars}. Notice that there are other instances of such antisymmetric couplings, namely the di-quark couplings of $\omega_4$, $\zeta$, and $\Omega_1$. Since in those cases also other couplings (with no particular symmetry structure) are present, we set the antisymmetric ones to zero\footnote{Notice that this choice doesn't change the sensitivity to EWPOs, cf. Tab. \ref{tab:scalars}}.
\end{itemize}   
In the conventions of Ref.~\cite{deBlas:2017xtg} we set all (non-gauge) dimensionless couplings to unity, in natural units, for the sake of simplicity of presentation. For all trilinear scalar couplings we choose the scale $\kappa = 5$ TeV, in order to avoid the additional suppression in $\kappa/M$ contributions to the dimension-six coefficients. Note that perturbativity is preserved in those cases, given the values of the masses we obtain. We do not include any interactions involving the new states at mass dimension five or above. 
We find that this would not affect our qualitative conclusions, since these couplings never change whether a state contributes at tree or one loop level to EWPOs.

We calculate the pattern of Wilson coefficients at the EW scale by running the full set of tree-level operators from the matching scale down to the EW scale, thus capturing the leading-log IR-calculable contributions to the EWPOs. Note that for consistency this is done also in the cases for which contributions to EWPOs arise already at tree level.\footnote{In these cases we also show pure tree-level bounds without RG effects, as hatched bars on the constraint plots. We find that inclusion of RG effects in these cases leads to a slight modification of the bounds, with a few exceptions for which one-loop effects dominate (see detailed discussion below).}  Running is calculated at the first leading-log level and is, for indicative purposes, run from a scale of $2$ TeV down to the EW scale using {\tt DsixTools} \cite{Celis:2017hod,Fuentes-Martin:2020zaz}. This scale has been chosen to roughly correspond to the average bounds that we find on states contributing to EWPOs via RGEs only. 

This RG evolution leads, in almost all cases, to non-vanishing contributions to the operators responsible for EWPOs.  Unless otherwise stated the one-loop matching contributions from within the UV theory are not included.  We opt for this choice for a number of reasons.  The first is that we wish to answer the question of whether a UV completion can generate dimension-six SMEFT contributions without contributing to EWPOs at the one-loop level.  Since we can, in almost all cases, answer this question in the affirmative already with IR-calculable effects it is not necessary to appeal further to one-loop matching terms.  The second reason is that these contributions are very model-dependent, sensitive to a much wider range of BSM couplings than the tree-level contributions, hence assigning too much importance to their numerical value when working in the couplings $ \to 1$ assumption would be taking the models too seriously at the quantitative level, especially when it is qualitative lessons we aim to extract.  Note that for one-loop matching one opens up sensitivity to all couplings which preserve a discrete $\mathcal{Z}_2$ symmetry acting on the new states, unlike for the tree-level terms.  Finally, we are not aware of any automated package capable of calculating the full set of matching coefficients in the Warsaw basis for all of the models considered in this work \footnote{Although automated software packages for one-loop matching such as {\tt Matchete}~\cite{Fuentes-Martin:2022jrf} and {\tt Matchmakereft}~\cite{Carmona:2021xtq} exist, they are currently not able to treat the integration of vector states consistently, which is why we refrain from using them in this work.}, without which such an extensive programme of calculations would be prohibitively onerous considering the minor quantitative and negligible qualitative impact on the results. The only exception is for the models $\Theta_1$ and $\Theta_3$, for which the leading contributions to EWPOs arise at the one-loop matching level and are included here, for the sake of demonstrating that they, too, do not evade EWPOs at one-loop.

\subsubsection*{Scalars}
The mass ranges probed for scalar UV completions are depicted in Fig.~\ref{tab:scalars}.  Only the scalar weak triplets $\Xi$ and $\Xi_1$ always generate tree-level EWPO contributions. In this case they violate custodial symmetry and thus the projected bounds are strong, well above the $10$ TeV level.  As previously discussed, neither $\Theta_1$ nor $\Theta_3$ generate tree-level or RG contributions, however one-loop custodial-violating contributions arise at the matching scale, also leading to strong constraints.

\begin{figure}[t]
    \centering
    \includegraphics[width=0.85\textwidth]{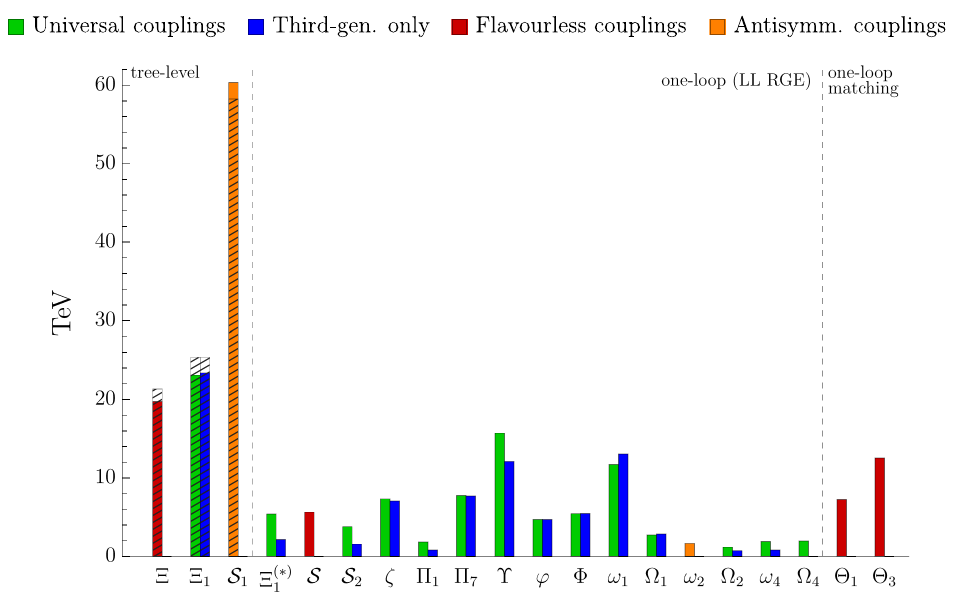}
    \label{fig:newscalars}
    \caption{Projected bounds (95\% CL) on the masses of new scalar fields. The vertical dashed lines separate fields which contribute to EWPOs at tree level (left), via one-loop RG evolution (middle), and via one-loop matching (right). The green and blue bars correspond to different assumptions for the coupling to SM fermions, as described near the start of Sec.~\ref{sec:fit}, while red bars are used when the state does not have couplings to fermions. Fields indicated with a $^{(*)}$ correspond to cases where the tree-level contribution has been set to zero by forbidding a specific coupling (see Table \ref{tab:scalars}). Hatched bars correspond to pure tree-level limits, without RG running.}
\end{figure}

The QCD sextuplet $\Omega_4$ also presents an interesting exception, generating only $\cC_{uu}$ at tree-level.  This feeds into $\cC_{Hu}$ under RG-evolution, giving rise to non-vanishing, yet relatively weak projected constraints.  On the other hand, for the third-generation only scenario only $\cC_{tt}$, a four top-right interaction, is generated which does not contribute to EWPOs under leading RG evolution.  This interesting exception warrants further study and, at present, is the only exception across all scalar UV-completions which can generate a tree-level dimension-6 SMEFT contribution without also leading to EWPO corrections at tree or one-loop level. Note, however, that if one resums beyond the first leading-log level this operator does generate an RG contribution to $C_{HD}$ \cite{Allwicher:2023aql,Stefanek:2024kds}, hence this model, too, is in fact not immune to EWPO measurements. 

For the remainder of the states the mass ranges probed span approximately $1-10$ TeV for states whose contribution to EWPOs is only through one-loop RG contributions, showing the extraordinary power of a Tera-$Z$ programme in directly probing beyond the mass ranges accessible to the HL-LHC, especially for purely EW-charged states, even at the one-loop level.  In the majority of cases the third-generation only ansatz is more weakly probed than the universal one, with only three exceptions ($\Xi_1$, $\zeta$ and $\omega_1$).

These can be traced back to the structure of the SMEFT RGE, together with the particular contributions to different EWPOs in each case.
A generic feature that always appears in this context is that the leading top Yukawa running of four-fermion operators into $Z$ coupling modifications ($H\psi$ operators) typically has the opposite sign than the gauge coupling contribution. This often leads to a cancellation in the RGE which is stronger in the case of universal couplings due to the sum over all flavours, and is typically of ${\cal O}(10)\%$.
On top of that, the case of $\Xi_1$ can be explained by a cancellation in $\delta m_W$ from $\cC_{ll}$ in the universal case, while for the leptoquark $\omega_1$ the coupling structure makes sure that $\tau$ and top coupling modifications always go together, yielding sensitivity to different observables in the two scenarios (cf. Table \ref{tab:scalars}).

The states with the lowest projected sensitivities are $SU(2)_L$ singlets which couple only to quarks; $\omega_2$, $\Omega_2$, $\omega_4$ and $\Omega_4$.\footnote{The state $\Omega_1$ also fits this definition, but has stronger projected constraints since its gauge structure allows it to match to $C_{qq}^{(1,3)}$, which can then run into EWPOs with large $y_t^2$ and $g_2^2$ prefactors.} These are scalar diquarks which are either triplets or sextets of QCD, and are best targeted at hadron colliders or in quark flavour observables (see e.g.~\cite{Giudice:2011ak}).

\subsubsection*{Vectors}
\begin{figure}[t]
    \centering
    \includegraphics[width=0.85\textwidth]{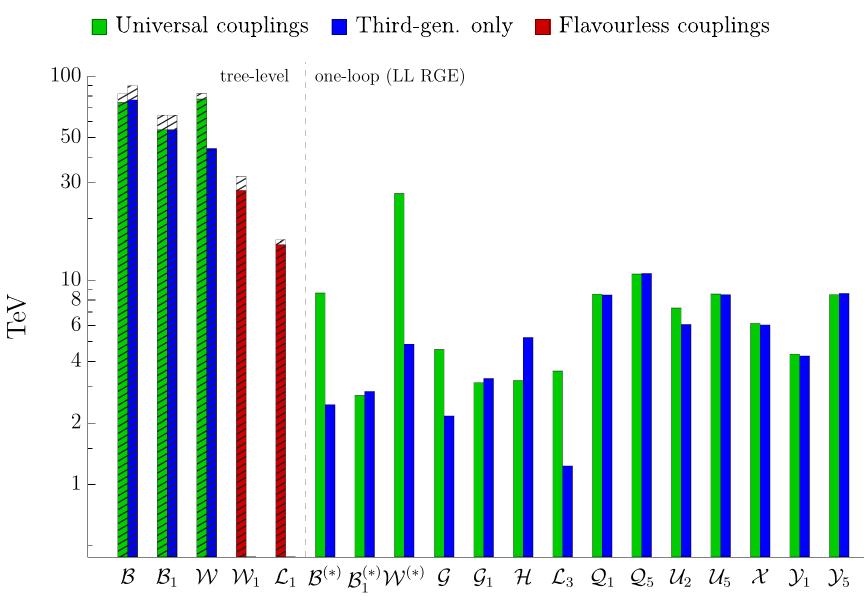}
    \caption{Projected bounds (95\% CL) on the masses of new vector fields. The vertical dashed line separates fields which contribute to EWPOs at tree level (left) and via one-loop RG evolution (right). The green and blue bars correspond to different assumptions for the coupling to SM fermions, as described near the start of Sec.~\ref{sec:fit}, while red bars are used when the state does not have couplings to fermions. Fields indicated with a $^{(*)}$ correspond to cases where the tree-level contribution has been set to zero by forbidding a specific coupling (see Table \ref{tab:vectors}). Hatched bars correspond to pure tree-level limits, without RG running.}
        \label{fig:newvectors}
\end{figure}

The mass ranges probed for vector UV completions are depicted in Fig.~\ref{fig:newvectors}, ranging over $1.5-80$ TeV.  Five scenarios ($\mathcal{B},~\mathcal{B}_1, \mathcal{W},~\mathcal{W}_1,~\mathcal{L}_1$) generically give rise to tree-level EWPO contributions if they generate any dimension-six operators at tree-level. For these, the projected constraints from the Tera-$Z$ programme at FCC-ee are at the $20$ TeV level and beyond.  However, for $\mathcal{B},~\mathcal{B}_1$ and $\mathcal{W}$ there are non-generic but plausible exceptions detailed in Tab.~\ref{tab:vectors} for which the tree-level contributions are absent.  In this case EWPO sensitivity is generated at one-loop under RG evolution, leading to the weaker constraints associated with the asterisked models in Fig.~\ref{fig:newvectors}.

For $\mathcal{B}$, $\mathcal{W}$ and $\mathcal{G}$ there are further conditions under which the one-loop RG contributions also vanish.  For $\mathcal{G}$ this can arise for only $(g^u_\mathcal{G})_{33} \neq 0$ since in that case only $\cC_{tt}$ is generated, which does not feed into EWPO under RG evolution. Note, however, that similarly to the $\Omega_4$ case, evolution beyond first leading-log does generate a contribution to $C_{HD}$ \cite{Allwicher:2023aql,Stefanek:2024kds}, which would lead to EWPO sensitivity. For $\mathcal{B}$ and $\mathcal{W}$ the exceptions at one loop of Tab.~\ref{tab:vectors} persist, however as discussed in Sec.~\ref{sec:exceptions}, the required charge assignments can lead to tree-level FCNCs, hence they may evade EWPOs, but not flavour constraints.

For the other scenarios bounds all exceed the TeV scale, in some cases by a large margin, and are stronger for the universal coupling ansatz, as expected, except for the cases of $\mathcal{G}_1$ and $\mathcal{H}$ where the projected bounds are mildly stronger in the third-generation only scenario. 
For $\mathcal{G}_1$ this is again due to cancellations in the RGE, while for $\cal H$ this can be traced back to a different sensitivity of the observables $R_\mu$ and $R_b$ to the $Z \bar q q$ coupling modifications. Together with the structure of the RGE, this leads to a suppression of the NP effect in $R_\mu$ for universal couplings.
All other cases with almost vanishing differences between the two flavour scenarios ($\cal B$, ${\cal Q}_5$, and ${\cal Y}_5$) 
are due to the fact that the leading constraint does not depend on the flavour assumption, or to larger cancellations in the RG evolution in the universal case, as above.

All-in-all we see that if perturbatively coupled vectors in the UV give rise to dimension-six SMEFT operators at tree level then they will also give contributions that enter into EWPOs at one loop at the least. 

Similarly to the case of the scalars, the weakest projected bounds are on states which couple only to quarks; ${\cal B}_1^{(*)}$, $\cal G$, ${\cal G}_1$, $\cal H$, ${\cal Y}_1$ and ${\cal Y}_3$. All of these except the special case ${\cal B}_1^{(*)}$ carry colour charge, and are targets for hadron colliders and precision quark flavour, demonstrating the complementarity between lepton and hadron machines.

\subsubsection*{Fermions}
Finally we come to the fermions.  As shown in Tab.~\ref{tab:fermions} for generic couplings all fermions which generate dimension-six tree-level contributions to the SMEFT also give EWPO contributions and are thus very effectively probed by an EWPO programme at FCC-ee.

\begin{figure}[t]
    \centering
    \includegraphics[width=0.85\textwidth]{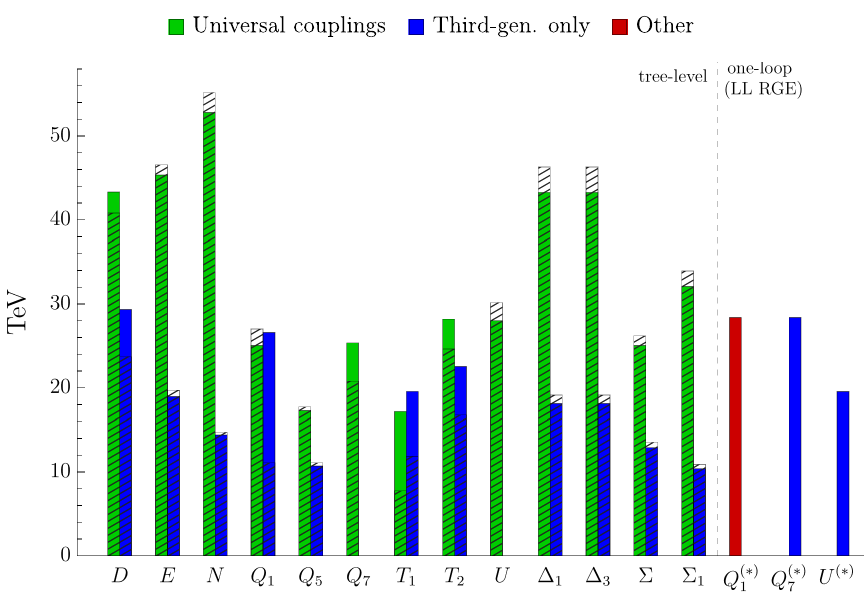}
    \caption{Projected bounds (95\% CL) on the masses of new fermion fields. The vertical dashed line separates fields which contribute to EWPOs at tree level (left) and via one-loop RG evolution (right). The green and blue bars correspond to different assumptions for the coupling to SM fermions, as described near the start of Sec.~\ref{sec:fit}, while the red bar for $Q_1^(*)$ corresponds to the exceptional case where $Q_1$ couples only to right-handed top quarks, as noted in Table \ref{tab:fermions}. Fields indicated with a $^{(*)}$ correspond to cases where the tree-level contribution has been set to zero by forbidding a specific coupling (see Table \ref{tab:fermions}). Hatched bars correspond to pure tree-level limits, without RG running.}
    \label{fig:newfermions}    
\end{figure}

This is reflected in Fig.~\ref{fig:newfermions} where all projected bounds for the case of universal couplings exceed the $10$ TeV scale.  For third-generation only couplings the projected limits are a little weaker in general, falling below $10$ TeV for $N$ and $Q_5$, and disappearing entirely for $Q_7$ and $U$.  For $U$ the reason is non-trivial.  In this model $\cC^{(1)}_{H{q_3}}$ and $\cC^{(3)}_{H{q_3}}$ are both generated at tree-level and both contribute to EWPOs.  However, they arise with equal and opposite Wilson coefficients which is precisely the condition of Eq.~\eqref{eq:universal1} in which EWPO contributions cancel, thus sensitivity is deferred to the one-loop level.  For $Q_7$ the reason is that only $C_{H{t}}$ is generated at tree-level in the third-generation only scenario, with no EWPO contributions, but ultimately running into $\cC_{HD}$ at one-loop.  Similarly, for $Q_1$ if only $(\lambda^u_{Q_1})_3\neq 0$ then only $\cC_{H{t}}$ is generated at tree-level with the same consequences as for $Q_7$. Note however how the bounds coming from the purely one-loop cases are essentially the same in size as for the tree-level EWPO cases. This is due to the fact that, given the large contribution from the top Yukawa in the running, $m_W$ is dominating the bounds in all cases. This feature is also highlighted by the hatched bars in Figure \ref{fig:newfermions}, showing significantly weaker bounds when RG effects are neglected.

In summary, fermionic perturbative UV completions which give rise to tree-level EWPO contributions are comprehensively probed by an EWPO programme at FCC-ee.  The exceptional cases give rise to EWPO contributions at the one-loop level and are still constrained beyond the $10$ TeV scale for $\mathcal{O}(1)$ couplings.

\section{Conclusion}
\label{sec:concs}
Through an unprecedented combination of precision with energy the FCC Tera-$Z$ programme promises to deliver the most comprehensive indirect probe of particle physics beyond the Standard Model at the highest energy scales hitherto accessible.  As yet the scope of this programme is poorly understood, often misunderstood.  It does not just promise an improved bound on the $S$ and $T$ parameters.  Rather, the precision proffers to probe disparate sectors of the SM by taking advantage of their quantum inseparability. 

Motivated and somewhat inspired by these early studies we have, in this work, attempted to examine this phenomenon from a broader, less model-inspired perspective.  We find that for the vast majority of perturbative UV completions which can generate a dimension-six contribution to the SMEFT at tree-level then a Tera-$Z$ EWPO programme has sensitivity at least at the one-loop level.  Extreme precision plays counterpoint to loop factors, allowing Tera-$Z$ to probe well above the TeV scale, in some cases up to $10$'s of TeV scales for completions with $\mathcal{O}(1)$ couplings.

We have performed fits to projected Tera-$Z$ measurements for each of the full set of states which match to the SMEFT at dimension six~\cite{deBlas:2017xtg}, under simple flavour assumptions on their fermionic couplings. As a prelude to this, we have also qualitatively studied the general parameter space of each state, to try to find exceptions or cancellations which avoid contributions to EWPOs at tree level and/or one loop. Remarkably, we find very few of these, demonstrating that Tera-$Z$'s sensitivity is not easily avoided by symmetries or flavour assumptions, unless tree level matching to the dimension six SMEFT is eliminated altogether. 

Tera-$Z$ will open a new precision frontier, but of course any UV model can also be tested by other datasets, including direct searches at the LHC, flavour measurements at LHCb,  Belle-II, and Tera-$Z$ itself, and measurements of Higgs and top properties during other runs of FCC-ee. In this work we do not discuss these complementary datasets, which will in some cases have better sensitivity to particular UV states. The message of our work is instead that EWPOs at Tera-$Z$ provide sensitivity to essentially \emph{any} weakly coupled new physics at the TeV scale or better. This message does not detract from the fact that in order to narrow in on any signals of new physics, or to find the best bound on individual models, different observables will need to be used in combination. Our analysis adds to the growing consensus that any such combination will need to include renormalisation effects, due to the separation between new physics and the electroweak scale, as well as the hierarchies of precision among measurements that will only deepend in the FCC-ee era.

The advent of an FCC-ee era would mark a quantum leap in our ability to broadly explore physics at the most microscopic scales.  So, too, must our theoretical framework for addressing questions that seek beyond the Standard Model enter a new paradigm, wherein the SMEFT RG-evolution \cite{Jenkins:2013zja,Jenkins:2013wua,Alonso:2013hga} is sure to play a significant and central role throughout an era of such extraordinary precision.

\acknowledgments
We are grateful to Jorge de Blas, Joe Davighi, Admir Greljo, Christophe Grojean, Gino Isidori, Ben Stefanek and Tevong You for discussions.
SR is supported by UKRI Stephen Hawking Fellowship EP/W005433/1, and partially supported by STFC grant ST/X000605/1. The work of LA is funded by the Swiss National Science Foundation (SNF) under contract 200020\_204428.

\appendix

\section{EWPO operators generated by each state}\label{ap:ops}

\begin{table}
  \centering
  \begin{tabular}{|c|l|l|}
  \hline
    Scalar & Tree Operators & One-Loop RG-Generated \\
    \hline
    $\mathcal{S}$ &
    $\mathcal{O}_{H\Box}$, [$\mathcal{O}_{H B}$],
    [$\mathcal{O}_{H W}$] & ($\mathcal{O}^{(1)}_{H q}$, $\mathcal{O}^{(3)}_{H q}$, $\mathcal{O}^{(1)}_{H l}$,  $\mathcal{O}^{(3)}_{H l}$, $\mathcal{O}_{H e}$)$^{g^2,y^2}$, \\
    & & ($\mathcal{O}_{H d}$, $\mathcal{O}_{H u}$)$^{g^2,y^2}$, ($\mathcal{O}_{H D}$,$\mathcal{O}_{H WB}$)$^{g^2}$  \\
    $\textcolor{red}{\mathcal{S}_1}$ &
    $\textcolor{red}{\mathcal{O}_{ll}}$ &  \\
    $\mathcal{S}_2$ &
    $\mathcal{O}_{ee}$  &  $\mathcal{O}_{H e}^{g^2,y^2}$ \\
    $\varphi$ &
    $\mathcal{O}_{le}$, $\mathcal{O}^{(1)}_{qu}$,
    $\mathcal{O}^{(1)}_{qd}$ & ($\mathcal{O}^{(1)}_{H q}$, $\mathcal{O}^{(1)}_{H l}$,  $\mathcal{O}_{H e}$, $\mathcal{O}_{H d}$, $\mathcal{O}_{H u}$)$^{g^2,y^2}$  \\
    $\textcolor{red}{\Xi}$ &
    $\textcolor{red}{\mathcal{O}_{H D}}$, $\mathcal{O}_{H\Box}$,
    $[\textcolor{red}{\mathcal{O}_{H WB}}$] &   \\
    $\textcolor{red}{\Xi_1}$ & 
    $\textcolor{red}{\mathcal{O}_{ll}}$, 
    $\textcolor{red}{\mathcal{O}_{H D}}$, $\mathcal{O}_{H\Box}$ &   \\
    $\Theta_1$ & $-$ & $\textcolor{blue}{\mathcal{O}_{H D}}$    \\
    $\Theta_3$ &  $-$ & $\textcolor{blue}{\mathcal{O}_{H D}}$   \\
    $\omega_{1}$ & 
    $\mathcal{O}^{(1)}_{qq}$, $\mathcal{O}^{(3)}_{qq}$,
    $\mathcal{O}^{(1)}_{lq}$, $\mathcal{O}^{(3)}_{lq}$, &  ($\mathcal{O}^{(1)}_{H q}$, $\mathcal{O}^{(3)}_{H q}$, $\mathcal{O}^{(1)}_{H l}$, $\mathcal{O}^{(3)}_{H l}$, $\mathcal{O}_{H e}$, $\mathcal{O}_{H d}$, $\mathcal{O}_{H u}$)$^{g^2,y^2}$, \\
    & $\mathcal{O}_{eu}$, $\mathcal{O}^{(1)}_{ud}$ &  $\mathcal{O}^{g^2}_{ll}$ \\
    $\omega_{2}$ & 
    $\mathcal{O}_{dd}$ & $\mathcal{O}_{H d}^{g^2,y^2}$  \\
    $\omega_{4}$ & 
    $\mathcal{O}_{uu}$, $\mathcal{O}_{ed}$ & ($\mathcal{O}_{H d}$, $\mathcal{O}_{H u}$, $\mathcal{O}_{H e}$)$^{g^2,y^2}$  \\
    $\Pi_1$ & 
    $\mathcal{O}_{ld}$ & ($\mathcal{O}^{(1)}_{H l}$, $\mathcal{O}_{H d}$)$^{g^2,y^2}$  \\
    $\Pi_7$ & 
    $\mathcal{O}_{lu}$, $\mathcal{O}_{qe}$ &  ($\mathcal{O}^{(1)}_{H l}$, $\mathcal{O}_{H u}$ $\mathcal{O}^{(1)}_{H q}$, $\mathcal{O}_{H e}$)$^{g^2,y^2}$ \\
    $\zeta$ &
    $\mathcal{O}^{(1)}_{qq}$, $\mathcal{O}^{(3)}_{qq}$,
    $\mathcal{O}^{(1)}_{lq}$, $\mathcal{O}^{(3)}_{lq}$ & ($\mathcal{O}^{(1)}_{H q}$, $\mathcal{O}^{(3)}_{H q}$, $\mathcal{O}^{(1)}_{H l}$, $\mathcal{O}^{(3)}_{H l}$)$^{g^2,y^2}$, $\mathcal{O}^{g^2}_{ll}$  \\
    $\Omega_{1}$ & 
    $\mathcal{O}^{(1)}_{qq}$, $\mathcal{O}^{(3)}_{qq}$,
    $\mathcal{O}^{(1)}_{ud}$ &  ($\mathcal{O}^{(1)}_{H q}$, $\mathcal{O}^{(3)}_{H q}$, $\mathcal{O}_{H d}$, $\mathcal{O}_{H u}$)$^{g^2,y^2}$  \\
    $\Omega_{2}$ & 
    $\mathcal{O}_{dd}$ &  $\mathcal{O}_{H d}^{g^2,y^2}$ \\
    $\Omega_{4}$ & 
    $\mathcal{O}_{uu}$ &  $\mathcal{O}_{H u}^{g^2,y^2}$ \\
    $\Upsilon$ & 
    $\mathcal{O}^{(1)}_{qq}$, $\mathcal{O}^{(3)}_{qq}$ &  ($\mathcal{O}^{(1)}_{H q}$, $\mathcal{O}^{(3)}_{H q}$)$^{g^2,y^2}$ \\
    $\Phi$ &
    $\mathcal{O}^{(1)}_{qu}$,
    $\mathcal{O}^{(1)}_{qd}$  & ($\mathcal{O}^{(1)}_{H q}$, $\mathcal{O}_{H u}$, $\mathcal{O}_{H d}$)$^{g^2,y^2}$ \\
    \hline
  \end{tabular}
  \caption{Dimension-six operators contributing to EWPOs generated by the heavy scalar fields introduced in Ref.~\cite{deBlas:2017xtg}. Operators enclosed within square brackets are only generated via non-renormalisable interactions of the given scalar. Operators and models entering EWPOs at tree-level are shown in red.  Operators which upon RG evolution generate EWPO operators are shown in black.  In blue are operators generated at one-loop at the matching scale \cite{Durieux:2022hbu}.  The last column details the EWPO operators generated at one-loop through RG evolution.  For models already constrained at tree-level the one-loop contributions are not shown.  The superscript denotes the order in couplings $g$ for EW gauge, $y$ for the associated Yukawa coupling, at which the Wilson coefficient arises. }
  \label{tab:ops_scalars}
\end{table}

\newpage

\begin{table}
  \centering
  \begin{tabular}{|c|l|}
  \hline
    Fermion & Tree Operators  
    \\
        \hline
    $\textcolor{red}{N}$ & 
    $\textcolor{red}{\mathcal{O}^{(1)}_{H l}}$,
    $\textcolor{red}{\mathcal{O}^{(3)}_{H l}}$     \\
    $\textcolor{red}{E}$ & 
    $\textcolor{red}{\mathcal{O}^{(1)}_{H l}}$, $\textcolor{red}{\mathcal{O}^{(3)}_{H l}}$     \\
    $\textcolor{red}{\Delta_1}$ & 
    $\textcolor{red}{\mathcal{O}_{H e}}$     \\
    $\textcolor{red}{\Delta_3}$ & 
    $\textcolor{red}{\mathcal{O}_{H e}}$     \\
    $\textcolor{red}{\Sigma}$ & 
    $\textcolor{red}{\mathcal{O}^{(1)}_{H l}}$, $\textcolor{red}{\mathcal{O}^{(3)}_{H l}}$     \\
    $\textcolor{red}{\Sigma_1}$ & 
    $\textcolor{red}{\mathcal{O}^{(1)}_{H l}}$, $\textcolor{red}{\mathcal{O}^{(3)}_{H l}}$     \\
    $\textcolor{red}{U}$ & 
    [$\mathcal{O}_{uB}$],
    $\textcolor{red}{\mathcal{O}^{(1)}_{H q}}$,
    $\textcolor{red}{\mathcal{O}^{(3)}_{H q}}$    \\
    $\textcolor{red}{D}$ & 
    $\textcolor{red}{\mathcal{O}^{(1)}_{H q}}$,
    $\textcolor{red}{\mathcal{O}^{(3)}_{H q}}$     \\
    $\textcolor{red}{Q_1}$ & 
    [$\mathcal{O}_{uB}$],
    [$\mathcal{O}_{uW}$], 
    $\textcolor{red}{\mathcal{O}_{H d}}$, $\textcolor{red}{\mathcal{O}_{H u}}$     \\
    $\textcolor{red}{Q_5}$ & 
    $\textcolor{red}{\textcolor{red}{\mathcal{O}_{H d}}}$     \\
    $\textcolor{red}{Q_7}$ & 
    $\textcolor{red}{\mathcal{O}_{H u}}$     \\
    $\textcolor{red}{T_1}$ & 
    $\textcolor{red}{\mathcal{O}^{(1)}_{H q}}$,
    $\textcolor{red}{\mathcal{O}^{(3)}_{H q}}$     \\
    $\textcolor{red}{T_2}$ &
    [$\mathcal{O}_{uW}$], $\textcolor{red}{\mathcal{O}^{(1)}_{H q}}$,
    $\textcolor{red}{\mathcal{O}^{(3)}_{H q}}$    \\
\hline
  \end{tabular}
  \caption{Same as for Tab.~\ref{tab:ops_scalars}, however generated by the heavy vector-like fermions introduced in Ref.~\cite{deBlas:2017xtg}.}
  \label{tab:ops_fermions}
  \end{table}

\begin{table}
  \centering
  \begin{tabular}{|c|l|l|}
  \hline
    Vector & Tree Operators  & One-Loop RG-Generated
    \\
    \hline
    $\textcolor{red}{\mathcal{B}}$ &
    $\textcolor{red}{\mathcal{O}_{ll}}$, $\mathcal{O}^{(1)}_{qq}$,
    $\mathcal{O}^{(1)}_{lq}$, $\mathcal{O}_{ee}$,
    $\mathcal{O}_{dd}$, $\mathcal{O}_{uu}$,
    $\mathcal{O}_{ed}$, & \\
    &
    $\mathcal{O}_{eu}$,
    $\mathcal{O}^{(1)}_{ud}$, $\mathcal{O}_{le}$,
    $\mathcal{O}_{ld}$, $\mathcal{O}_{lu}$,
    $\mathcal{O}_{qe}$, $\mathcal{O}^{(1)}_{qu}$, & \\
    &
    $\mathcal{O}^{(1)}_{qd}$, $\textcolor{red}{\mathcal{O}_{H D}}$, $\mathcal{O}_{H \Box}$, $\textcolor{red}{\mathcal{O}^{(1)}_{H l}}$, & \\
    &  $\textcolor{red}{\mathcal{O}_{H u}}$,
    $\textcolor{red}{\mathcal{O}^{(1)}_{H q}}$, $\textcolor{red}{\mathcal{O}_{H e}}$,
    $\textcolor{red}{\textcolor{red}{\mathcal{O}_{H d}}}$ &    \\
    $\textcolor{red}{\mathcal{B}_1}$ &
    $\mathcal{O}^{(1)}_{ud}$,
    $\textcolor{red}{\mathcal{O}_{H D}}$, $\mathcal{O}_{H \Box}$ &    \\
    $\textcolor{red}{\mathcal{W}}$ &
    $\textcolor{red}{\mathcal{O}_{ll}}$,
    $\mathcal{O}^{(3)}_{qq}$, $\mathcal{O}^{(3)}_{lq}$,
    $\textcolor{red}{\mathcal{O}_{H D}}$,
    $\mathcal{O}_{H \Box}$, $\textcolor{red}{\mathcal{O}^{(3)}_{H l}}$, $\textcolor{red}{\mathcal{O}^{(3)}_{H q}}$ &    \\
    $\textcolor{red}{\mathcal{W}_1}$ & 
    $\textcolor{red}{\mathcal{O}_{H D}}$, $\mathcal{O}_{H \Box}$ &    \\
    $\mathcal{G}$ & 
    $\mathcal{O}^{(1)}_{qq}$, $\mathcal{O}^{(3)}_{qq}$,
    $\mathcal{O}_{dd}$, $\mathcal{O}_{uu}$ & ($\mathcal{O}^{(1)}_{H q}$, $\mathcal{O}^{(3)}_{H q}$,$\mathcal{O}_{H d}$, $\mathcal{O}_{H u}$)$^{g^2,y^2}$   \\
    $\mathcal{G}_1$ & 
    $\mathcal{O}^{(1)}_{ud}$ &  ($\mathcal{O}_{H d}$, $\mathcal{O}_{H u}$)$^{g^2,y^2}$  \\
    $\mathcal{H}$ & 
    $\mathcal{O}^{(1)}_{qq}$, $\mathcal{O}^{(3)}_{qq}$ & ($\mathcal{O}^{(1)}_{H q}$, $\mathcal{O}^{(3)}_{H q}$)$^{g^2,y^2}$  \\
    $\textcolor{red}{\mathcal{L}_1}$ &
    $\mathcal{O}^{(1)}_{qu}$,
    $\mathcal{O}^{(1)}_{qd}$, $\textcolor{red}{\mathcal{O}_{H D}}$,
    $\mathcal{O}_{H \Box}$, $\mathcal{O}_{H B}$,
    $\mathcal{O}_{H W}$, & \\
    &
    $\mathcal{O}_{H WB}$, 
    $\mathcal{O}_{uB}$, $\mathcal{O}_{uW}$,
    $\textcolor{red}{\mathcal{O}^{(1)}_{H l}}$, $\textcolor{red}{\mathcal{O}^{(3)}_{H l}}$,
    $\textcolor{red}{\mathcal{O}^{(1)}_{H q}}$, &  \\
    &
    $\textcolor{red}{\mathcal{O}^{(3)}_{H q}}$, $\textcolor{red}{\mathcal{O}_{H e}}$, $\textcolor{red}{\textcolor{red}{\mathcal{O}_{H d}}}$,
    $\textcolor{red}{\mathcal{O}_{H u}}$  &   \\
    $\mathcal{L}_3$ & 
    $\mathcal{O}_{le}$ & ($\mathcal{O}^{(1)}_{H l}$, $\mathcal{O}_{H e}$)$^{g^2,y^2}$    \\
    $\mathcal{U}_2$ &
    $\mathcal{O}^{(1)}_{lq}$, $\mathcal{O}^{(3)}_{lq}$,
    $\mathcal{O}_{ed}$ & ($\mathcal{O}^{(1)}_{H q}$, $\mathcal{O}^{(3)}_{H q}$, $\mathcal{O}^{(1)}_{H l}$, $\mathcal{O}^{(3)}_{H l}$)$^{g^2,y^2}$, \\
    & & ($\mathcal{O}_{H e}$, $\mathcal{O}_{H d}$)$^{g^2,y^2}$, $\mathcal{O}^{g^2}_{ll}$  \\
    $\mathcal{U}_5$ & 
    $\mathcal{O}_{eu}$ & ($\mathcal{O}_{H e}$, $\mathcal{O}_{H u}$)$^{g^2,y^2}$  \\
    $\mathcal{Q}_1$ & 
    $\mathcal{O}_{lu}$, $\mathcal{O}^{(1)}_{qd}$ & ($\mathcal{O}^{(1)}_{H q}$, $\mathcal{O}_{H d}$, $\mathcal{O}^{(1)}_{H l}$, $\mathcal{O}_{H u}$)$^{g^2,y^2}$  \\
    $\mathcal{Q}_5$ & 
    $\mathcal{O}_{qe}$, $\mathcal{O}^{(1)}_{qu}$ & ($\mathcal{O}^{(1)}_{H q}$, $\mathcal{O}_{H e}$, $\mathcal{O}_{H u}$)$^{g^2,y^2}$  \\
    $\mathcal{X}$ &
    $\mathcal{O}^{(1)}_{lq}$ & ($\mathcal{O}^{(1)}_{H q}$, $\mathcal{O}^{(1)}_{H l}$)$^{g^2,y^2}$  \\
    $\mathcal{Y}_1$ & 
    $\mathcal{O}^{(1)}_{qd}$ & ($\mathcal{O}^{(1)}_{H q}$, $\mathcal{O}_{H d}$)$^{g^2,y^2}$   \\
    $\mathcal{Y}_5$ &
    $\mathcal{O}^{(1)}_{qu}$ & ($\mathcal{O}^{(1)}_{H q}$, $\mathcal{O}_{H u}$)$^{g^2,y^2}$   \\
\hline

  \end{tabular}
  \caption{Same as for Tab.~\ref{tab:ops_scalars}, however generated by the heavy vector bosons introduced in Ref.~\cite{deBlas:2017xtg}.}
  \label{tab:ops_vectors}
\end{table}

\bibliographystyle{JHEP}
\bibliography{refs}

\end{document}